\documentclass[10pt]{aastex}
\usepackage{emulateapj5}
\usepackage{graphicx}
\def\be{\begin{equation}}
\def\ee{\end{equation}}

\def\be{\begin{equation}}
\def\ee{\end{equation}}
\catcode`\@=11 % This allows us to modify PLAIN macros.
\def\@versim#1#2{\vcenter{\offinterlineskip
        \ialign{$\m@th#1\hfil##\hfil$\crcr#2\crcr\sim\crcr } }}
\def\lsim{\mathrel{\mathpalette\@versim<}}
\def\gsim{\mathrel{\mathpalette\@versim>}}
\def\mpy{M_\odot \ {\rm yr^{-1}}}
\def\kms{{\rm km \, s^{-1}}}

\shorttitle{}
\shortauthors{}

\begin{document}
%\date{}

%\title{High Energy $\gamma$-rays from the Galactic Center}
\title{Nonthermal THz to TeV Emission from Stellar Wind Shocks in the
Galactic Center}

\author{Eliot Quataert\altaffilmark{1} and Abraham Loeb\altaffilmark{2}}

\email{eliot@astron.berkeley.edu; aloeb@cfa.harvard.edu}

\altaffiltext{1}{Astronomy Department, 601 Campbell Hall, University of
California, Berkeley, CA 94720}

\altaffiltext{2}{Harvard-Smithsonian Center for Astrophysics, 60 Garden
Street, Cambridge, MA 02138}

\begin{abstract}

The central parsec of the Galaxy contains dozens of massive stars with
a cumulative mass loss rate $\sim 10^{-3} \, \mpy$.  Shocks among
these stellar winds produce the hot plasma that pervades the central
part of the galaxy.  We argue that these stellar wind shocks also
efficiently accelerate electrons and protons to relativistic energies.
The relativistic electrons inverse Compton scatter the ambient
ultraviolet and far infrared radiation field, producing high energy
$\gamma$-rays with a roughly constant luminosity from $\sim$ GeV to
$\sim 10$ TeV.  This can account for the TeV source seen by HESS in
the Galactic Center.  Our model predicts a {\it GLAST} counterpart to
the HESS source with a luminosity of $\approx 10^{35}$ ergs s$^{-1}$
and cooling break at $\approx 4$ GeV.  Synchrotron radiation from the
same relativistic electrons should produce detectable emission at
lower energies, with a surface brightness $\approx 10^{32} B^2_{-3}$
ergs s$^{-1}$ arcsec$^{-2}$ from $\sim$ THz to $\sim$ keV, where
$B_{-3}$ is the magnetic field strength in units of mG. The observed
level of diffuse thermal X-ray emission in the central parsec requires
$B \lsim 300 \, \mu$ G in our models.  Future detection of the diffuse
synchrotron background in the central parsec can directly constrain
the magnetic field strength, providing an important boundary condition
for models of accretion onto Sgr A*.

\end{abstract}

\section{Introduction}

The young stellar cluster in the central parsec of the Galaxy contains
nearly two dozen luminous blue supergiants and Wolf-Rayet stars, in
addition to a larger population of lower mass O and B stars (Krabbe et
al. 1995; Genzel et al. 2003).  These stars were produced in a short lived
compact starburst that occurred $\sim 10^6-10^7$ years ago.  Collectively,
the massive stars in the central parsec lose $\approx 10^{-3} \, \mpy$ in
stellar winds with velocities ranging from $\sim 300$ to $1000 \, \kms$
(Najarro et al. 1997; Paumard et al. 2001).  Shocks produced by colliding
stellar winds can account for the hot ($\approx 1-2$ keV) X-ray emitting
plasma that fills the central few parsecs of the Galaxy (Quataert 2004;
Rockefeller et al. 2004; Cuadra et al. 2005).  These stellar winds are also
the most likely source of fuel for the central massive black hole Sgr A*:
$\approx 0.3-1 \%$ of the mass lost to stellar winds is gravitationally
captured by the black hole and flows in to smaller radii (e.g., Coker \&
Melia 1997; Quataert 2004; Cuadra et al. 2005).  The remaining mass is
thermally driven out of the central stellar cluster in a wind.

In this {\it Letter} we investigate the possibility that stellar wind
shocks in the Galactic center also efficiently accelerate electrons and
protons to ultra-relativistic energies.  This claim is motivated by the
analogy between the strong collisionless shocks expected in the Galactic
center and similar shocks in supernovae remnants and $\gamma$-ray bursts
that are inferred to be efficient sites of particle acceleration (e.g.,
Drury 1983; Blandford \& Eichler 1987). We show that a population of
relativistic particles in the central parsec of the galaxy produces
detectable emission from the radio to the $\gamma$-rays (\S2).  We then
compare the predictions of our model with current observations and discuss
its further implications (\S3).  We focus primarily on emission by
relativistic electrons because it has fewer uncertainties, but we also
briefly consider pion production by relativistic protons.

\section{Predicted Spectrum}

The total kinetic energy thermalized by shocks in the Galactic center
is $\dot E = {1 \over 2} f {\dot M_w} v_w^2 \approx 10^{38} f_{0.3}
{\dot M_{-3}} v_8^2$ ergs s$^{-1}$, where $\dot M_w = 10^{-3} \dot
M_{-3} \, \mpy$ is the total stellar mass loss rate, $v_w = 1000 \, v_8
\, \kms$ is the average wind speed, and $f = 0.3 f_{0.3}$ accounts for
geometric uncertainties such as the fraction of the stellar wind
material that undergoes strong shocks (the normalization for $f$ is
taken from the 3D simulations of Rockefeller et al. 2004).

We assume that a fraction $\eta = 10^{-2} \eta_{-2}$ of the shock
energy goes into accelerating relativistic electrons, where the
normalization of $\eta$ is motivated by observations of supernova
remnants (see, e.g., \S 2.2.2 of Keshet et al. 2003 for a summary of
the relevant observations).  Thus the energy supplied to relativistic
electrons is \be \dot E_e \approx 10^{36} f_{0.3} \eta_{-2} {\dot
M_{-3}} v_8^2 \, {\rm ergs \, s^{-1}}. \label{electron} \ee We assume
for simplicity that the electrons are injected with a spectrum
$\propto \gamma^{-p}$ with $p = 2$, similar to the index typically
observed in supernova remnants (e.g., Brogan et al. 2005; Aharonian et
al. 2005) and theoretically expected for strong shocks (Bell 1978;
Blandford \& Ostriker 1978).

The high energy electrons cool by synchrotron radiation and inverse Compton
(IC) emission.  Figure 1 summarizes our predictions for the spectrum
produced by these two processes; the main features in this spectrum are
described analytically below.  The magnetic field in the central parsec is
poorly constrained.  We normalize our estimates using $B = B_{-3}$ mG,
which is equipartition with the hot thermal plasma and is similar to the
field strength inferred using the properties of non-thermal radio filaments
on larger scales in the Galactic center (e.g., Morris \& Serabyn 1996; see,
however, LaRosa et al. 2005).  Equipartition field strength could in
principle also be generated by the shocks themselves (e.g., Medvedev \&
Loeb 1999).

The young stellar cluster in the central parsec has an ultraviolet (UV)
luminosity of $L = 10^{41} L_{41}$ ergs s$^{-1}$ within a radius $R=0.5
R_{0.5}$ pc (Krabbe et al. 1995), implying a radiation energy density of
$U_{\rm ph} = 10^{-7} U^{\rm UV}_{-7}$ ergs cm$^{-3}$ with a characteristic
photon energy of $E_{\rm ph} \approx 10$ eV.  Only $\approx 20 \%$ of the UV
radiation is absorbed by dust {\it in situ} (Davidson et al. 1992), which
leads to a far infrared (FIR) radiation field with $E_{\rm ph} \approx 0.04$ eV
and $U^{\rm IR}_{-7} \sim 0.1$ or equivalently $U^{\rm IR}_{-8}\sim 1$.

The relativistic particles in the Galactic center are advected out of
the central parsec with the thermally driven wind on a timescale
$t_{exp} \approx R/v_w \approx 500 \, R_{0.5}/v_8$ years.  Sufficiently
relativistic electrons with $\gamma > \gamma_{c}$ cool on a timescale
$\lsim t_{exp}$.  For our fiducial normalization the cooling is
primarily by IC radiation (unless $B \gsim 1.6$ mG). In this case the
critical Lorentz factor where the IC cooling time equals the expansion
time is given by \be \gamma_c \approx {3 \pi m_e c^2 R v_w \over
\sigma_T L} \approx 2 \times 10^4 {R_{0.5} v_8 \over L_{41}}
\label{gcool} \ee where $\sigma_T$ is the Thompson cross section.
Electrons with $\gamma > \gamma_c$ lose all of their energy to IC
radiation before escaping the central parsec, producing a power \be
\nu L_\nu|_{IC} \approx {\dot E_e \over \ln(\gamma_{max})} \approx 6
\times 10^{34} f_{0.3} \eta_{-2} {\dot M_{-3}} v_8^2 \, {\rm ergs \,
s^{-1}}
\label{spect} 
\ee where $\gamma_{max}$ is the maximum Lorentz factor of the accelerated
electrons.  The exact value of $\gamma_{max}$ is uncertain but a rough
estimate can be obtained by balancing the cooling time with the
acceleration time at the shock front, $t_{acc} \approx \Omega^{-1}
c^2/v_w^2$, where $\Omega$ is the relativistic cyclotron frequency (e.g.,
Loeb \& Waxman 2000).  This yields \be \gamma_{max} \sim \left(e B \over
\sigma_T U_{\rm ph}\right)^{1/2} {v_w \over c} \sim 3 \times 10^7
{B_{-3}^{1/2} v_8 \over \sqrt{U^{\rm IR}_{-8}}}. \label{gmax} \ee Note that
in the above numerical estimate we have normalized the photon energy
density using the FIR radiation field, while in our estimate of $\gamma_c$
in equation (\ref{gcool}) we used the full UV radiation field.  This is
because electrons with $\gamma \sim \gamma_{max}$ see UV photons in their
rest-frame as $\gamma$-rays with energy $\gg m_ec^2$ and do not efficiently
scatter the UV photons due to the Klein-Nishina suppression of the cross
section.

%Note also that the above
%expression for $\gamma_{max}$ assumes that the Thompson cross section
%is applicable which may not be true at the highest Lorentz factors.

%For simplicity, let's consider first the predicted spectrum for a
%monoenergetic source of seed photons in the UV; the effects of both IR
%and UV photons will be considered below.  

Equations (\ref{gcool}) and (\ref{gmax}) imply that the IC power in
equation (\ref{spect}) applies for photon energies $E_{min} < E <
E_{max}$, where
%Our basic prediction is that IC scattering leads to a high energy
%$\gamma$-ray spectrum with
%\be \nu L_\nu|_{IC} \approx {\dot E_e \over \ln(\gamma_{max})} \approx 2
%\times 10^{35} \eta_{-2} {\dot M_{-3}} v_8^2 \, {\rm ergs \, s^{-1}}
%\label{spect} 
%\ee
%for photon energies 
\be E_{min} \approx \gamma_c^2 E_{\rm ph} \approx 4 \, {\rm GeV} \left(R_{0.5}
v_8 \over L_{41}\right)^2 \left(E_{\rm ph} \over 10 \, eV\right) \label{Emin}
\ee and \be E_{max} \approx \gamma_{max}^2 E_{\rm ph} \approx 30 \, {\rm TeV}
\left(B_{-3} v_8^2 \over U^{\rm IR}_{-8}\right) \left(E_{\rm ph} \over 0.04 \,
eV\right). \label{Emax} \ee Relativistic corrections to the electron
scattering cross section lead to an additional suppression of the IC power
above \be E_b \approx 25 \, {\rm TeV} \, \left(E_{\rm ph} \over 0.04 \, {\rm
eV} \right)^{-1}.
\label{knbreak} \ee 
%Note that the maximum photon energy set by
%Klein-Nishina corrections (eq. [\ref{knbreak}]) is comparable to that
%set by the maximum Lorentz factor of the accelerated electrons
%(eq. [\ref{gmax}]).

For $E < E_{min}$, the electrons do not lose all of their energy
before flowing out of the central parsec and so the IC power is
suppressed, with $\nu L_\nu \propto (E/E_{min})^{1/2}$.  

\subsection{Synchrotron Emission}

The same electrons that produce IC power in the $\gamma$-rays will also
produce synchrotron emission at lower energies, with $P_s/P_{IC} =
U_B/U_{\rm ph}$.  For $\nu_c < \nu < \nu_{max}$ the total synchrotron
power is (assuming $U_B < U_{\rm ph}$) \be \nu L_\nu|_s \approx 2 \times
10^{34} \left(B_{-3}^2 \over U^{\rm UV}_{-7}\right) f_{0.3} \eta_{-2} {\dot
M_{-3}} v_8^2 \, {\rm ergs \, s^{-1}}
\label{synch} \ee where 
\be \nu_c \approx { \gamma_c^2 e B \over 2\pi m_e c} \approx 1 \, B_{-3}
\left(R_{0.5} v_8 \over L_{41}\right)^2 \, {\rm THz} \label{numin} \ee and
\be \nu_{max} \approx {\gamma_{max}^2 e B \over 2\pi m_e c} \approx 3
\times 10^{18} \left(B_{-3}^2 v_8^2 \over U^{\rm IR}_{-8}\right) \, {\rm
Hz}.
\label{numax} \ee  Near $\nu_{max}$ the electrons only IC 
scatter FIR photons and so $U_{ph} \approx U^{\rm IR}_{-8}$ should be
used in equation (\ref{synch}).\footnote{Note that if $U^{\rm IR}_{-8}
\sim 1$ and $B_{-3} \sim 1$, $U_B > U_{\rm ph}$ and so equation
(\ref{synch}) would not apply.  As we discuss in the next section,
however, observations constrain $B \lsim 300 \mu$G so this possibility
is not realized.}

Equations (\ref{synch})-(\ref{numax}) imply that synchrotron emission from
shock accelerated electrons produces a diffuse radiation field from the
radio to the X-rays in the central parsec.  This emission would be extended
on a radial scale of $\approx 0.5 R_{0.5} \, {\rm pc} \approx 10''$ and so
would have a surface brightness of \be I_s \approx 7 \times 10^{31}
\left(B_{-3}^2 \over L_{41} \right) f_{0.3} \eta_{-2} {\dot M_{-3}} v_8^2
\, {\rm ergs \, s^{-1} \, arcsec^{-2}}. \label{synchsb} \ee

\subsection{$\gamma$-ray Emission from Pion Decay}

Relativistic protons can also be a significant source of high energy
emission through the production of neutral and charged pions; here we
briefly estimate the expected $\gamma$-ray luminosity from neutral
pion production in the central few parsecs of the Galaxy.  We assume
that a larger fraction, $\sim 10 \%$, of the shock energy goes into
accelerating relativistic protons (Blandford \& Eichler 1987), in
which case \be \dot E_p \approx 10^{37} f_{0.3} \eta_{-1} {\dot
M_{-3}} v_8^2 \, {\rm ergs \, s^{-1}}. \label{proton} \ee The hot
thermal plasma in the Galactic center has a density of $n \approx 30$
cm$^{-3}$ (e.g., Baganoff et al. 2003).  The timescale for
relativistic protons to lose their energy via neutral pion production
is $t_{pion} \approx 10^8 \, n^{-1}$ years.  Comparing this to the
expansion time of the hot plasma $\sim 500$ years, we find a very low
GeV $\gamma$-ray luminosity from pion production on the ambient hot
plasma, $\nu L_\nu \approx 10^{-4} \dot E_p/\ln(\gamma_{max}) \sim
10^{32} \, {\rm ergs \, s^{-1}}$.

A more promising source of $\gamma$-rays from pion decay in the central
few parsecs is the interaction between the shock accelerated protons
and the circum-nuclear disk (CND).  The CND is a ring of dense
molecular gas located $\approx 1.5$ pc from Sgr A*, with a scale
height of $\sim 0.5$ pc (Jackson et al. 1993; Christopher et
al. 2005).  The mean density of the CND is at least $n \approx 10^5
n_5$ cm$^{-3}$, and Christopher et al. (2005) argue that it could be
several orders of magnitude higher.  The morphology of the diffuse
X-ray emission in the central parsec suggests that the outflowing wind
from the central star cluster impinges on and is confined by the CND
(Baganoff et al. 2003; Rockefeller et al. 2004).

To estimate the $\gamma$-ray luminosity from pion decay in the CND, we
assume that a fraction $\xi = 0.1 \xi_{0.1}$ of the relativistic
protons created in the central $0.5$ pc enter into the CND, and that a
fraction $\sim 1/3$ of the proton energy goes into $\gamma$-rays,
rather than neutrinos or other particles.  The normalization for $\xi$
is based on the solid angle subtended by the inner edge of the CND,
but in principle $\xi$ could be much smaller if relativistic particles
are magnetically shielded out of the CND.  The cooling time for
relativistic protons due to pion production in the CND is $t_{pion}
\approx 10^3 \, n_5^{-1}$ years.  By contrast, the timescale for
relativistic protons to escape from the CND is uncertain, and depends
sensitively on the magnetic field geometry and cross-field diffusion.
Unless the effective diffusion velocity is $\gsim 1000 \, \kms$,
however, the protons will lose all of their energy in the CND because
the cooling time is shorter the escape time.  Under this assumption,
the $\gamma$-ray luminosity from the CND is \be \nu L_\nu \sim {\xi
\dot E_p \over 3 \ln(\gamma_{max})} \sim 2 \times 10^{34} f_{0.3}
\xi_{0.1} \eta_{-1} {\dot M_{-3}} v_8^2 \, {\rm ergs \,
s^{-1}}. \label{pion} \ee Equation (\ref{pion}) predicts the maximum
$\gamma$-ray luminosity not far below that produced by relativistic
electrons (eq. [\ref{spect}]), but the estimate is much less certain
because the relativistic protons might quickly leak out of the CND
along open magnetic field lines.  The spectrum of $\gamma$-rays from
pion decay would extend from $\approx 70$ MeV to many TeV, with the
maximum photon energy determined by the maximum energy of the
accelerated protons.

\section{Discussion}

Figure 1 summarizes our estimates of the IC and synchrotron radiation
expected from shock accelerated electrons in the central parsec.  We do not
include the predicted radiation from pion decay estimated in \S2.3 because
it is much less certain, though we reiterate that the $\gamma$-ray flux from
pion decay in the CND could be non-negligible.

Figure 1 also summarizes some of the data relevant to our predictions.  In
particular, IC radiation of the ambient FIR radiation field can account for
the TeV source seen by HESS (Aharonian et al. 2005) and CANGAROO (Tsuchiya
et al. 2004) towards the Galactic center.  In Figure 1 we have adjusted the
normalization of the IC component to account for the HESS luminosity by
setting $\eta \approx 5\times 10^{-3}$ in equation (\ref{spect}).  Our
model naturally accounts for the HESS spectrum and makes the strong
prediction that the {\it Gamma-Ray Large Area Space Telescope} ({\it
GLAST})\footnote{See http://glast.gsfc.nasa.gov} should observe a cooling
break at $\approx 4$ GeV.  The energy of the cooling break is very well
determined in our model (eq. [\ref{Emin}]) since it is set by the UV
radiation energy density and the expansion time of the thermal plasma, both
of which are observationally well constrained.

Atoyan \& Dermer (2004) also explain the TeV emission from the
Galactic center as IC scattering of FIR photons by relativistic
electrons (see Pohl 1997 and Melia et al. 1998 for related ideas in
the context of the {\it EGRET} source 3EG J1746-2851, which is now
believed to be well-offset from the central parsec; Pohl 2005). In
their model, however, the electrons are produced in the termination
shock of an outflow from Sgr A*.  On energetic grounds, the shocked
stellar winds investigated in this {\it Letter} may be a more
important source of particle acceleration: the accretion rate onto Sgr
A* is believed to be $\lsim 10^{-8} \mpy$ (e.g., Agol 2000; Quataert
\& Gruzinov 2000), implying a total accretion power $\approx 0.1 \dot
M c^2 \lsim 5 \times 10^{37} \dot M_{-8} \, {\rm ergs \, s^{-1}}$,
somewhat smaller than the energy thermalized by shocked stellar winds
in the central parsec.  In addition, the presence of stellar wind
shocks in the central parsec is observationally well established,
while the properties of an outflow from Sgr A* (its mass loss rate,
collimation, etc.) are significantly less well understood.

In addition to IC emission at high energies, we also predict that
synchrotron radiation from shock accelerated electrons in the central
parsec leads to a diffuse background of radiation from $\sim$ THz to
$\sim$ keV (Fig. 1), with a surface brightness given by equation
(\ref{synchsb}). This emission would be extended on the scale of the
central stellar cluster, $\approx 10''$.  The magnitude of the diffuse
synchrotron emission is sensitive to the magnetic field strength in
the central parsec, which is poorly constrained
observationally.\footnote{On much larger ($\ga 2'$) scales, LaRosa et
al. (2005) argue that the field strength is $\sim 10\mu$G, while the
prevalence of nonthermal radio filaments has been used to argue for $B
\sim m$G (e.g., Morris \& Serabyn 1996).} The most stringent
observationally limits to date on diffuse emission at the relevant
wavelengths are probably from deep {\it Chandra} observations which
find extended thermal emission with a surface brightness of $\approx
7.6 \times 10^{31} \, {\rm ergs \, s^{-1} \, arcsec^{-2}}$ in the
central few parsecs (Baganoff et al. 2003).  Using equation
(\ref{synchsb}) with $L_{41} \approx 0.1$ appropriate for the FIR
radiation field scattered by electrons with $\gamma \sim
\gamma_{max}$, we find that the {\it Chandra} observations require
that the magnetic field strength in the central parsec be $\lsim 300
\mu G$ in order to not overproduce the extended X-ray emission.  It is
worth highlighting that future probes of the nonthermal diffuse
emission could directly constrain the magnetic field in the central
parsec.  In particular, the ratio of the luminosity of the diffuse
emission to the luminosity of the HESS TeV source is determined by
$U_B/U_{\rm ph}$ and thus can be used to measure $U_B$.  Such a
measurement would provide a very useful outer boundary condition for
models of accretion onto Sgr A*, which are sensitive to the geometry
and magnitude of the magnetic field at large radii (e.g., Igumenshchev
\& Narayan 2002).

Lastly, we note that some fraction of the shock accelerated electrons
and protons in the central parsec should be gravitationally captured
by Sgr A* along with the thermal plasma, providing a seed population
of relativistic particles to the accretion flow.  The Lorentz factor
of such particles will, at a minimum, increase via adiabatic
compression as $\gamma \propto \rho^{1/3}$.  The density contrast
between the Bondi accretion radius ($R_B \sim 10^5 R_s$) and the event
horizon is believed to be a factor of $\sim 10^5$ (see Quataert 2003
for a review), implying that mildly relativistic electrons at $\sim
R_B$ have $\gamma \sim 100$ close to the black hole.  Such electrons
would emit in the mm-MIR and could contribute significantly to the
observed emission from Sgr A*.  This possibility will be explored in
more detail in a future paper.

\acknowledgments We thank Reinhard Genzel and Chuck Dermer for useful
conversations.  EQ thanks the ITC at Harvard for their hospitality.  EQ is
supported in part by NSF grant AST 0206006, NASA grant NAG5-12043, an
Alfred P. Sloan Fellowship, and the David and Lucile Packard Foundation. AL
is supported in part by NASA grants NAG 5-13292 and NNG05GH54G.

\clearpage

\begin{figure}
\plotone{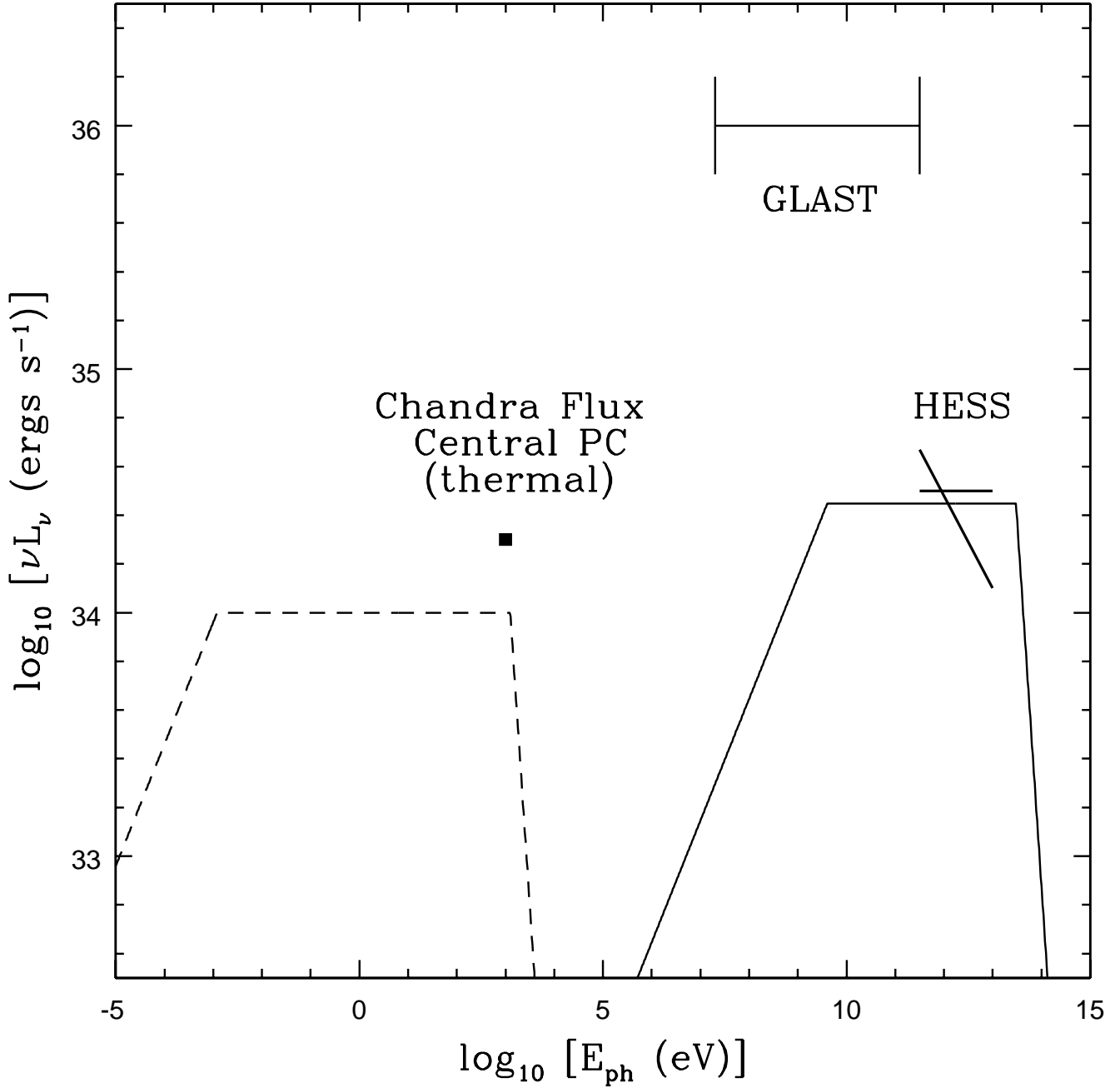}
\caption{Predicted emission from shock accelerated electrons in the
central parsec of the Galactic center.  The solid line shows the IC
radiation (eq. [\ref{spect}]) while the dashed line shows the
synchrotron radiation for $B = 0.3$ mG and $U_{\rm ph} = 10^{-8}$ erg
cm$^{-3}$ (eq. [\ref{synch}]).  The parameter $\eta$ measuring the
fraction of energy in shock accelerated electrons is adjusted to
reproduce the HESS TeV luminosity ($\eta \approx 5\times 10^{-3}$).
With this normalization the $\gamma$-ray spectrum is uniquely
determined, including the predicted cooling break in the {\it GLAST}
band at $\approx 4$ GeV; the flux of synchrotron radiation depends
primarily on the magnetic field strength in the central parsec.  The
high energy cutoffs to the IC and synchrotron radiation are determined
by $\gamma_{max}$, the maximum Lorentz factor of accelerated electrons
(eq. [\ref{gmax}]).  Klein-Nishina suppression of the electron
scattering cross section produces a high energy cutoff to the IC
emission at comparable energies (eq. [\ref{knbreak}]).  The exact
spectral shape above the high energy cutoff is uncertain and depends
on the details of of the electron distribution around $\gamma_{max}$.
The two power-laws shown for the HESS data indicate the uncertaintes
in the power-law slope (Aharonian et al. 2004).}
\end{figure}

\end{document}